\documentclass[twocolumn]{webofc}
\usepackage[varg]{txfonts}   
\usepackage{graphicx}
\usepackage{subcaption}
\usepackage{float} 

\begin{document}

\title{Preliminary results of the AMIGA engineering array at the Pierre Auger Observatory}
%
%

\author{\firstname{Alvaro} \lastname{Taboada}\inst{1,2}\fnsep\thanks{\email{alvaro.taboada@kit.edu}}, \thanks{\email{auger_spokespersons@fnal.gov}}  for the Pierre Auger Collaboration\inst{3}
}

\institute{Karlsruhe Institute of Technology, Karlsruhe, Germany
\and
          Instituto de Tecnologías en Detección y Astropartículas, CNEA-CONICET-UNSAM, Buenos Aires, Argentina
\and
           Pierre Auger Observatory, Malargüe, Argentina
         }

\abstract{%
The Auger Muons and Infill for the Ground Array (AMIGA) aims to both extend the detection range of the Pierre Auger Observatory
down to energies $\sim 10^{16.5}~\mathrm{eV}$ and to measure the muon content of extensive air showers.
To accomplish these goals, its detection system is composed of an array of coupled water-Cherenkov and scintillation detectors deployed in
a graded triangular grid of 433 and 750\,m spacings. 
At each position, the scintillation detector is buried $2.3~\mathrm{m}$ deep so as to shield it from the air shower electromagnetic component and thus only measure the muon component.
These muon detectors have $30~\mathrm{m^2}$ area split into modules, each of them highly segmented in 64 plastic-scintillator strips with an embedded wavelength-shifter optical fiber to transport light to an optical sensor located at the center of the module.
During the engineering array phase (finished in November 2017)
two module areas ($5~\mathrm{m^2}$ and $10~\mathrm{m^2}$) and two optical sensors (photo-multiplier tubes and silicon-photomultipliers) 
were tested. In this work, we present the final performance of the muon detectors equipped with silicon-photomultipliers which were thereafter selected as the baseline design for the AMIGA production phase. Analyses and results are based both on laboratory and field measurements.
}
\maketitle
\section{The AMIGA array}
\label{sec::intro}
The Pierre Auger Observatory completed its surface-detector array with $1500~\mathrm{m}$ spacing (SD-1500) in 2008 to study cosmic rays with energies above $10^{18.5}$ eV. An area of $3000~\mathrm{km^2}$ was fully deployed with water-Cherenkov detectors (WCD) \cite{Auger2015}. To extend the energy threshold down to $\sim 10^{16.5}$ eV, the Auger Muons and Infill for the Ground Array (AMIGA) \cite{AMIGA} proposed a smaller area of $23.5~ \mathrm{km^2}$ which has already been completely equipped with WCDs spaced at $750~\mathrm{m}$ (SD-750). More recently an even denser array at a $433~\mathrm{m}$ spacing with an area of $1.9~\mathrm{km^2}$ is being deployed (SD-433).
\begin{figure}[h]
\centering
\includegraphics[width=6.5cm,clip]{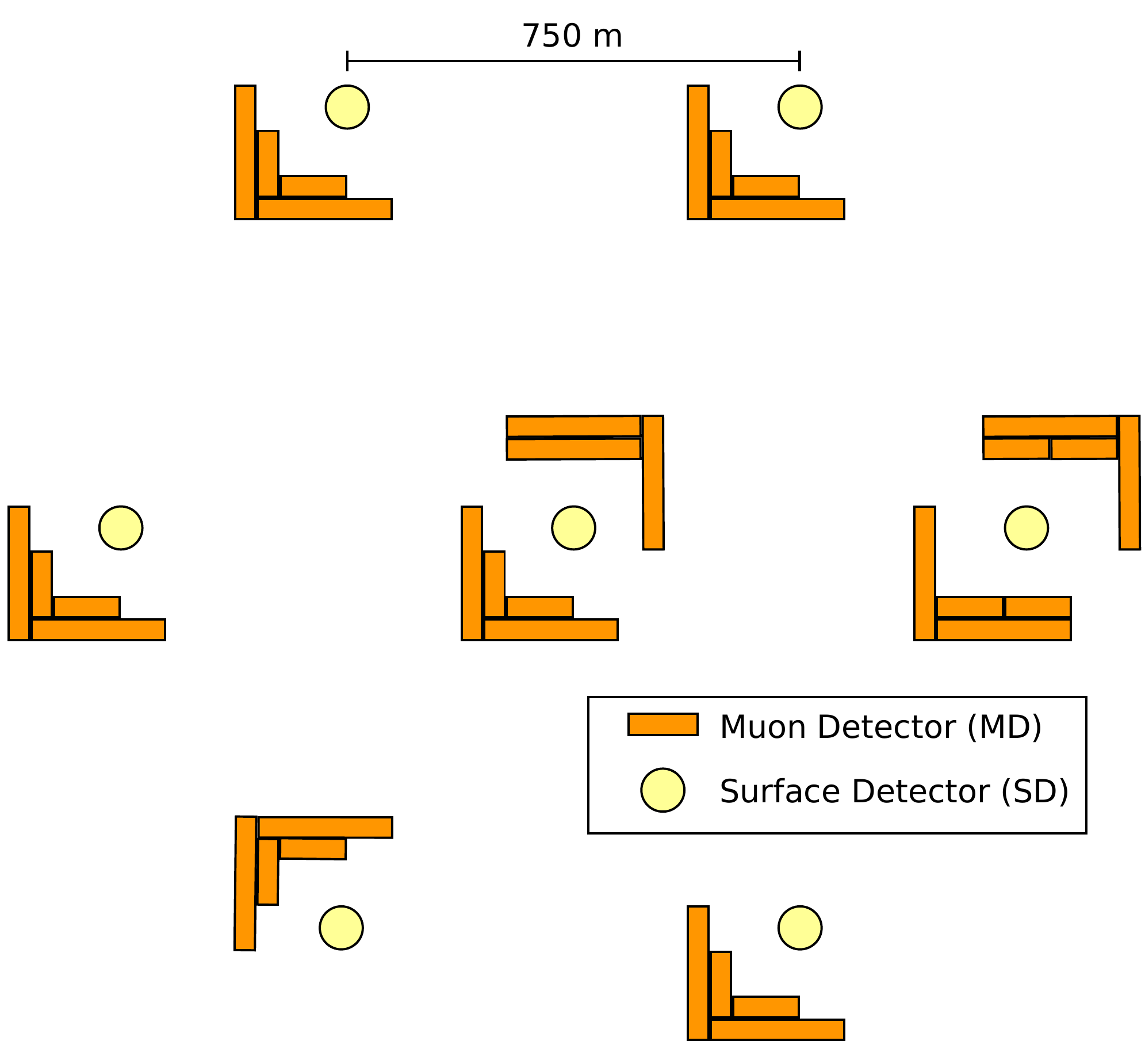}
\caption{Layout of the AMIGA EA. In this prototype phase the $30~\mathrm{m^2}$ UMDs were
	segmented in modules of $5~\mathrm{m^2}$ and $10~\mathrm{m^2}$. Also, two positions were equipped with extra {\it twin} modules
	to assess detector uncertainties. Two optical sensors were tested, multipixel  and silicon photomultipliers.}
\label{fig::mapEA}       
\end{figure}
Still, AMIGA's  main and novel contribution to the Pierre Auger Observatory is to provide direct muon measurements with dedicated underground muon detectors (UMD) made of plastic scintillators. Each UMD is coupled to a WCD of both SD-750 and SD-433 arrays. 
Between March 2015 and November 2017, an UMD engineering array (EA) was operated in the SD-750 as a proof of concept, for physics validation, and optimization of the detector design. The layout of this EA is shown in fig.~\ref{fig::mapEA}. 
\begin{figure*}[h]
\centering
\includegraphics[width=6.7cm,clip]{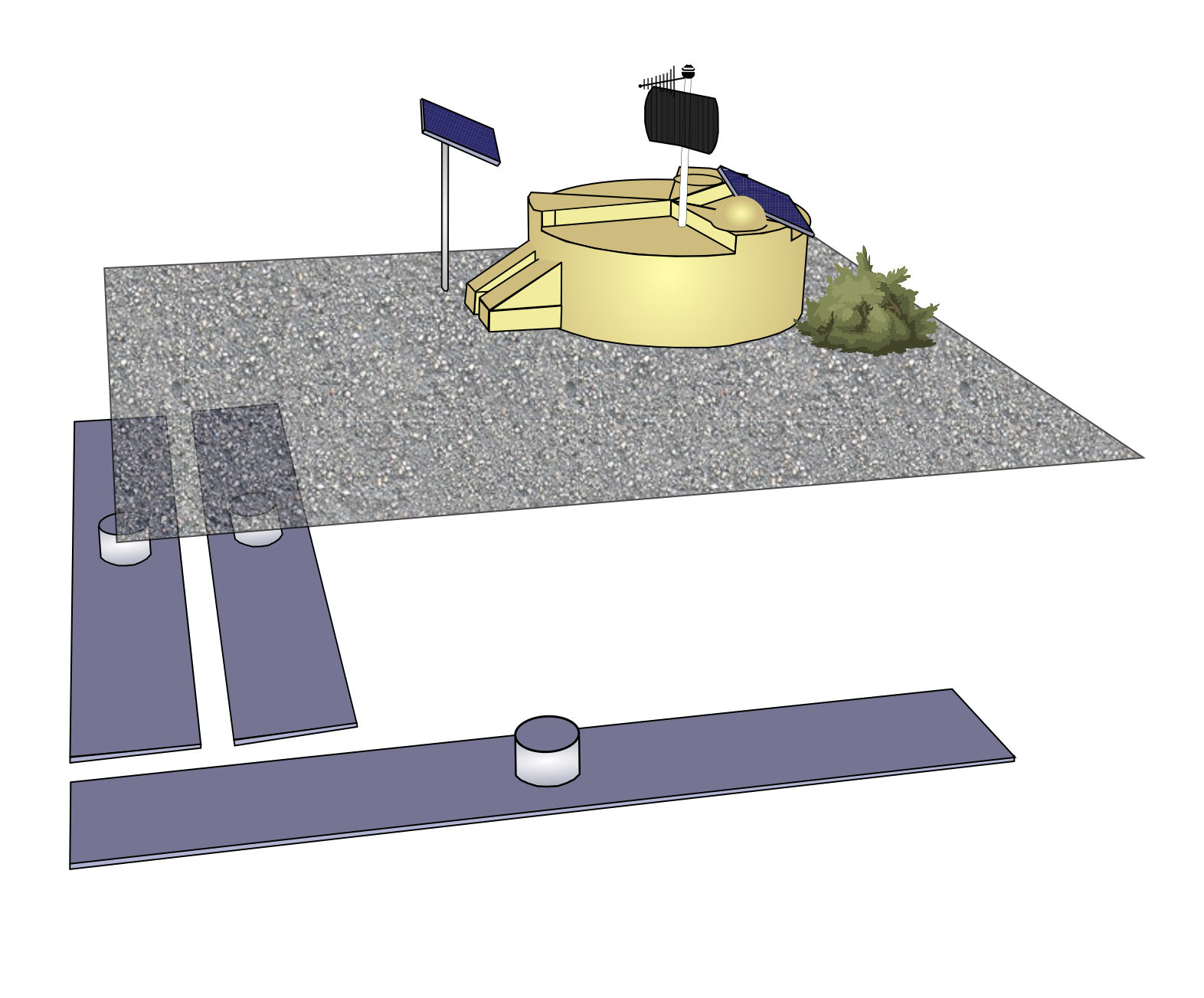}
\includegraphics[trim={18cm 18cm 0 18cm},width=4.2cm,clip]{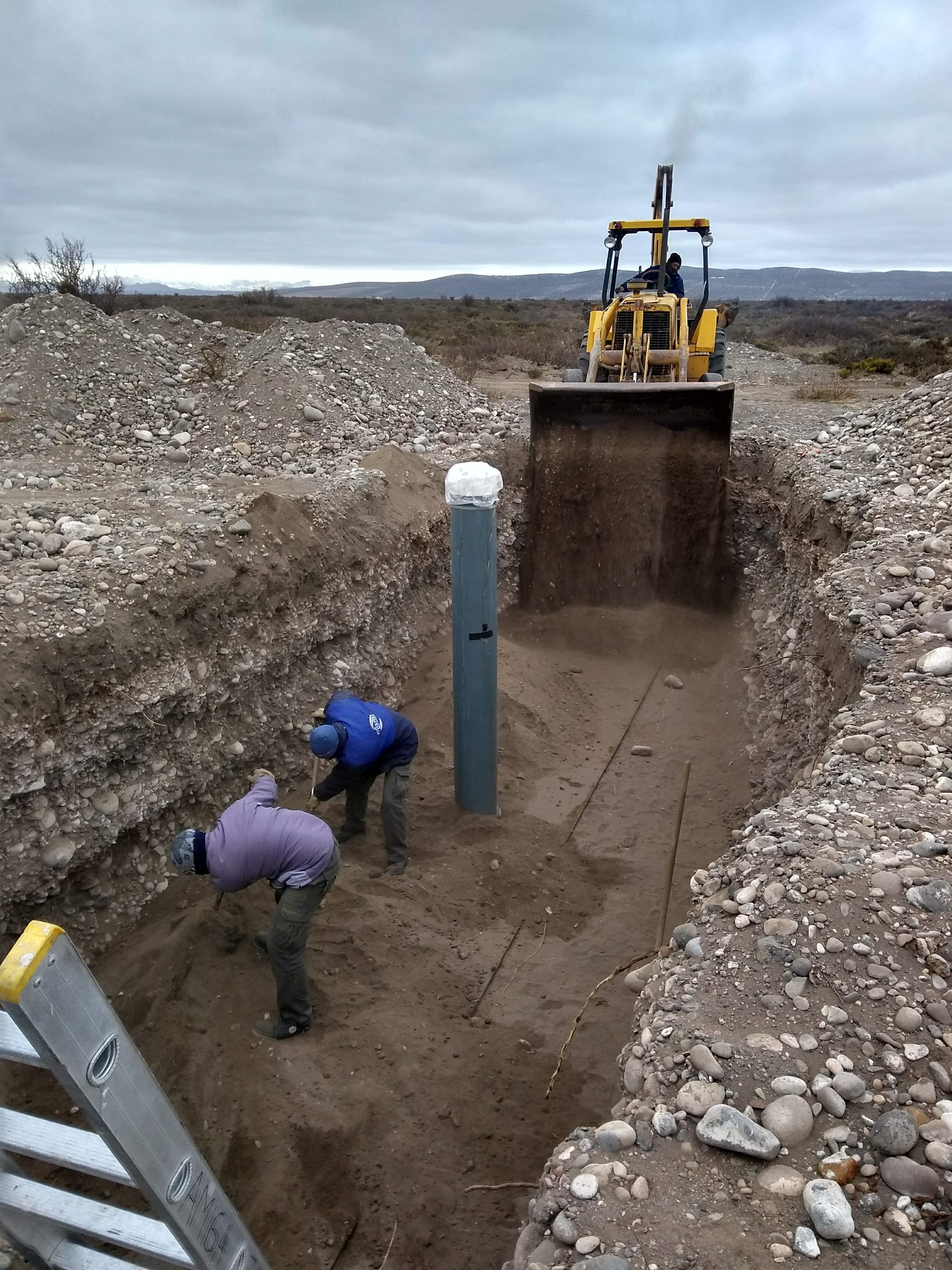}
	\caption{\textit{Left:} Final layout of the AMIGA production-phase design: each WCD is coupled to three $10~\mathrm{m^2}$ buried scintillator modules. \textit{Right:} Detector deployment in the field in which the tube that hosts the electronics and the optical sensor can be seen.}
\label{fig::MDSD}
\end{figure*}

\begin{figure*}[h]
\centering
\includegraphics[width=6.0cm,clip]{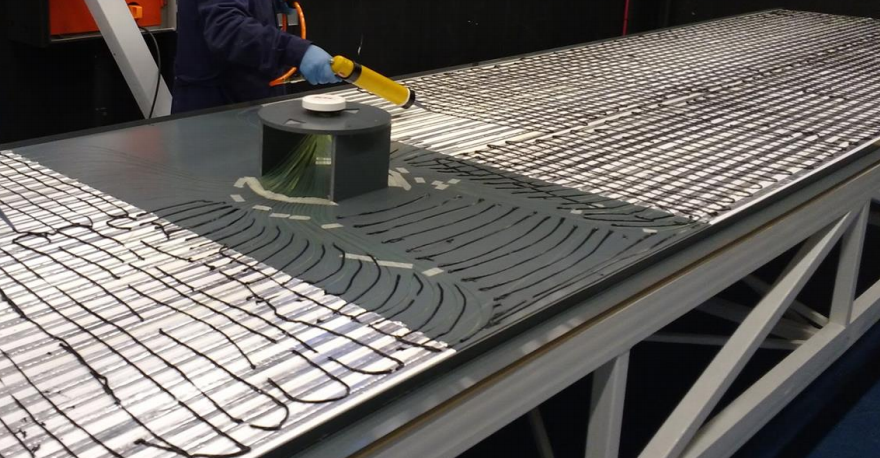}
\includegraphics[width=6.0cm,clip]{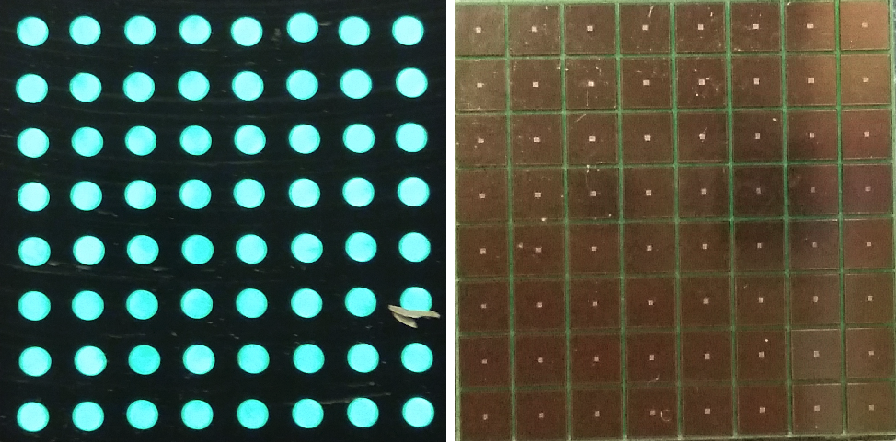}
	\caption{\textit{Left:} UMD fabrication and assembling. Each module is composed of 64 scintillator bars with a wavelength-shifter optical fiber glued to them. \textit{Middle:} Central $8 \times 8$ optical connector to which light is transported by the fibers. \textit{Right:} The SiPMs array which is coupled to the connector.}
\label{fig::sipmarray}       
\end{figure*}

The $30~\mathrm{m^2}$ of scintillators, buried $2.3~\mathrm{m}$ deep close to a WCD, were segmented in modules of 5 and 10 $\mathrm{m^2}$. In turns, each module consists of 64 extruded plastic-scintillator bars with a wavelength-shifter optical fiber glued
to transport light towards an optical connector located at the module center \cite{Suarez}. Two kinds of optical sensors were tested during the prototype phase: a 64 pixel multi-anode photomultiplier tube (MPMT) and an array of 64 silicon photomultipliers
(SiPM). After completion of these tests, the final decision for the production phase was to only build $10~\mathrm{m^2}$ modules equipped with SiPMs.
A schematic of the final layout is shown in fig.~\ref{fig::MDSD} alongside a field deployment in which is seen the inspection tube that hosts the electronics and the optical sensors.

In fig.~\ref{fig::sipmarray}, an example of a module fabrication and assembly is shown together with the chosen SiPM array (Hamamatsu S13361-2050NE-08) and the optical connector that couples the fibers onto it.

The electronics readout works simultaneously as a {\it counter} (counts number of arriving muons) and as an {\it integrator} (integrates the total signal of the arriving muons). 
The counter mode consists of 64 channels (one for each scintillator strip in a module) made of a pre-amplifier, a fast-shaper and, finally, a discriminator. These channels are integrated in two 32-pixel ASICs (Application-Specific Integrated Circuits). The signals after the discriminators are all identical in amplitude and only differ in their time width depending on the time the input pulses are above a given threshold.
All signals from the 64 discriminators are sampled by a Field-Programmable Gate Array (FPGA) with a sampling time of $3.125~\mathrm{ns}$ ($320~\mathrm{MHz}$). At each sampling, the FPGA associates a ``1'' if the discriminator pulse is present and a ``0'' otherwise. As a result, the electronics for the counter mode outputs 64 binary traces where individual muons, and their arrival time, can be easily measured by inspecting the strings of ``1s''. 
\begin{figure*}[h]
\centering
\includegraphics[width=5.5cm,clip]{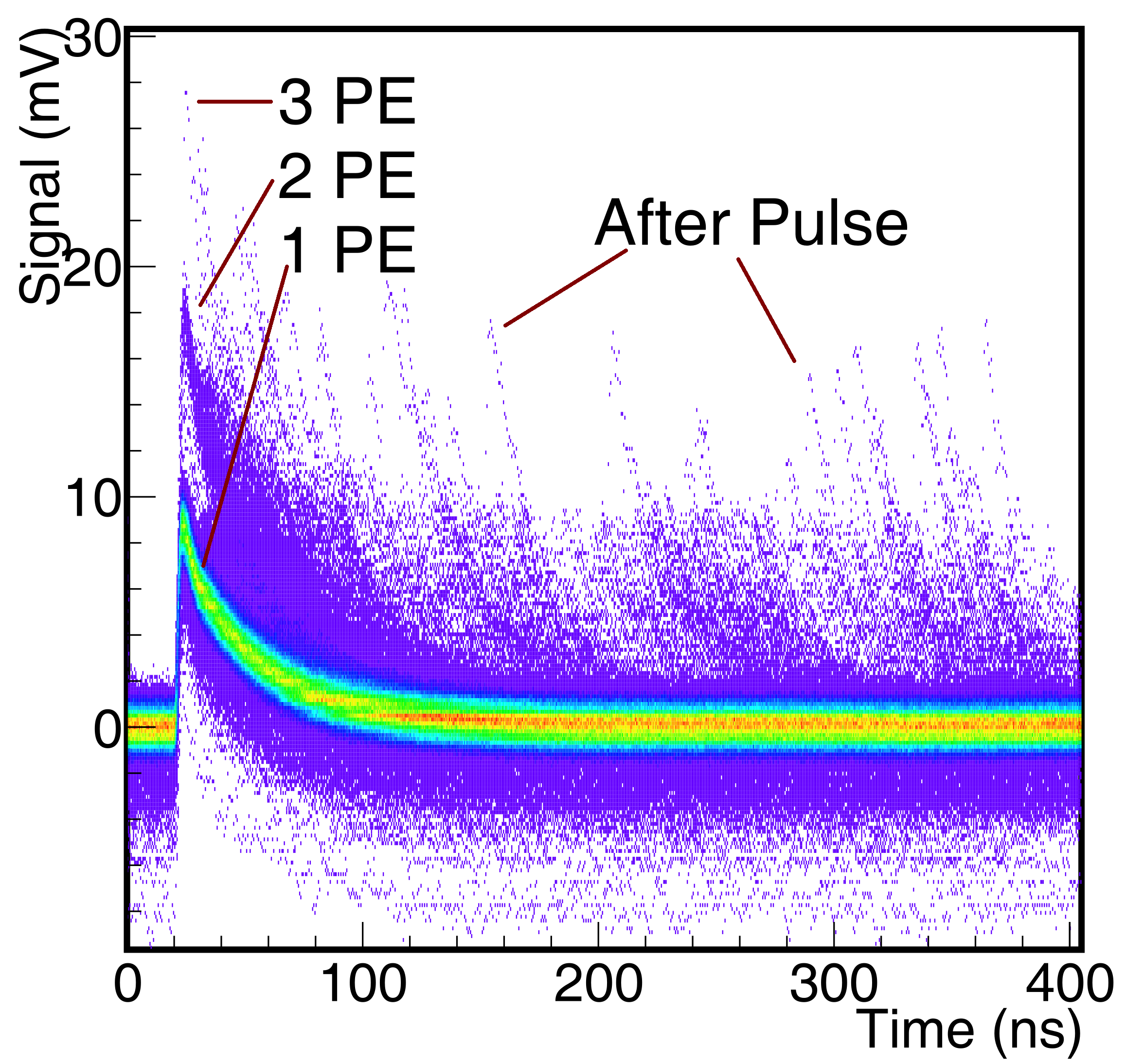}
\includegraphics[width=8.0cm,clip]{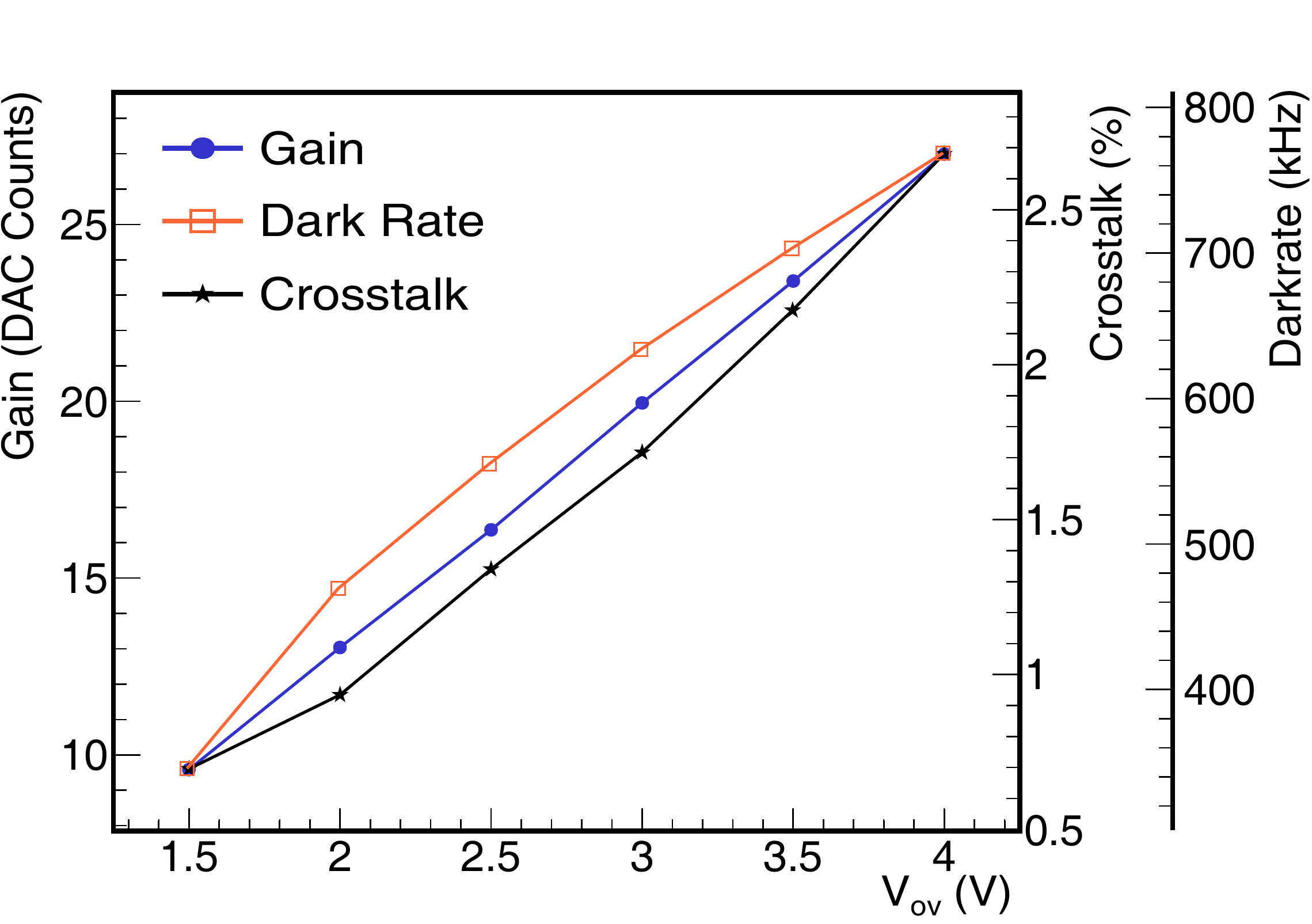}
	\caption{\textit{Left:} Oscilloscope capture of dark pulses from the AMIGA SiPM, crosstalk pulses and afterpulses are clearly visible. \textit{Right:} Gain, dark rate, and crosstalk as a function of the over-voltage.}
\label{fig::darkRate}      
\end{figure*}

\begin{figure*} [htp]
\centering
\includegraphics[width=17cm,clip]{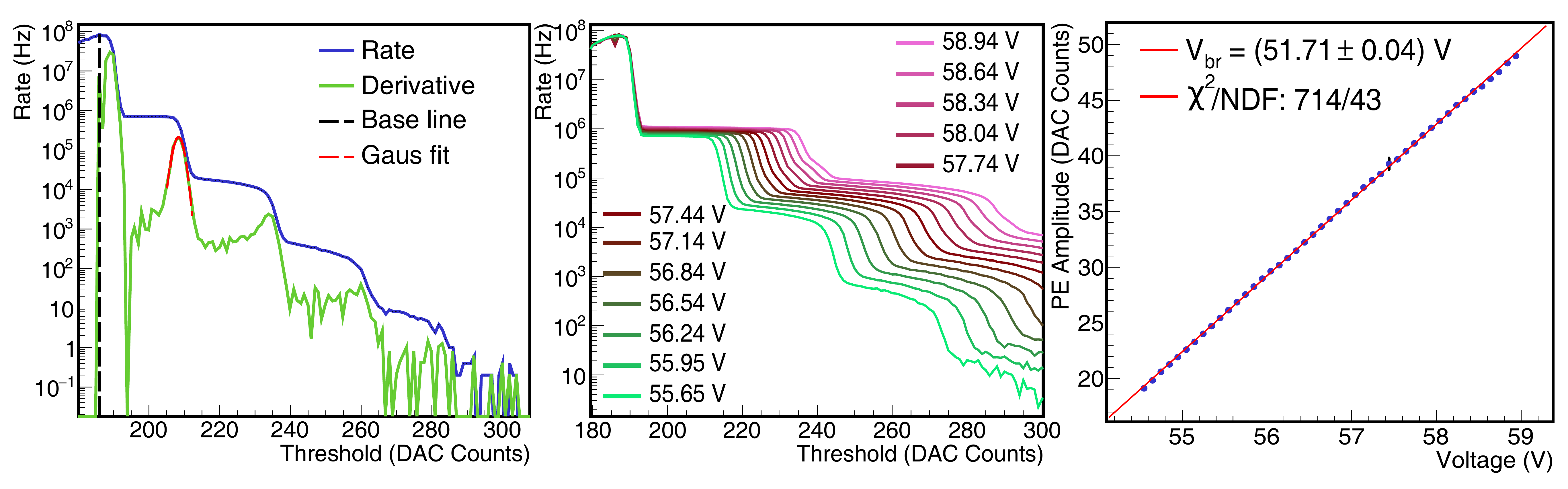}
	\caption{\textit{Left:} Dark-rate curve as a function of the discriminator threshold (blue). The first plateau (below 190 ADC counts) corresponds to the baseline while the followings indicate an increasing number of PE amplitudes. The derivative of this rate curve (green) readily identifies the PE amplitude spectrum. \textit{Middle:} Dark-rate curves at different $V_\mathrm{br}$. It is seen how the dark rates and PE amplitudes shift towards higher values when rising the $V_\mathrm{br}$. \textit{Right:} Determination of the breakdown voltage ($V_\mathrm{br}$) for an individual SiPM. The linear extrapolation to 0 amplitude indicates the $V_\mathrm{br}$ value.
	\label{fig::calibration}}
\end{figure*}

On the other hand, and complementary to the counter mode, the integrator mode of the front-end electronics adds up analogically the 64 SiPM signals and processes the result through a low-gain and a high-gain amplification channel. The amplified signals are sampled with Analog-to-Digital Converters (ADCs) with sample time of $6.25~\mathrm{ns}$ ($180~\mathrm{MHz}$), resulting in two waveforms of 1024 samples from which the total charge of the signal may be calculated. 

The counter mode is hardware limited by the segmentation of each module since if two muons arrive simultaneously in any given strip (i.e. within a preset time window \cite{AMIGA}, typically of 35 ns), they will pile up and be counted as a single muon\footnote{It is important to stress, that the time distribution of muons within a shower certainly allows one to measure more than one muon per strip in the counter mode, increasing noticeably the total number of non-simultaneous particles that can be measured}. The integrator mode solves this issue and extends the dynamic range of the detector to measure higher muon densities close to the shower core.

\section{SiPM operation}
\label{sec::sipms}
\begin{figure*}
\centering
\includegraphics[width=12.5cm,clip]{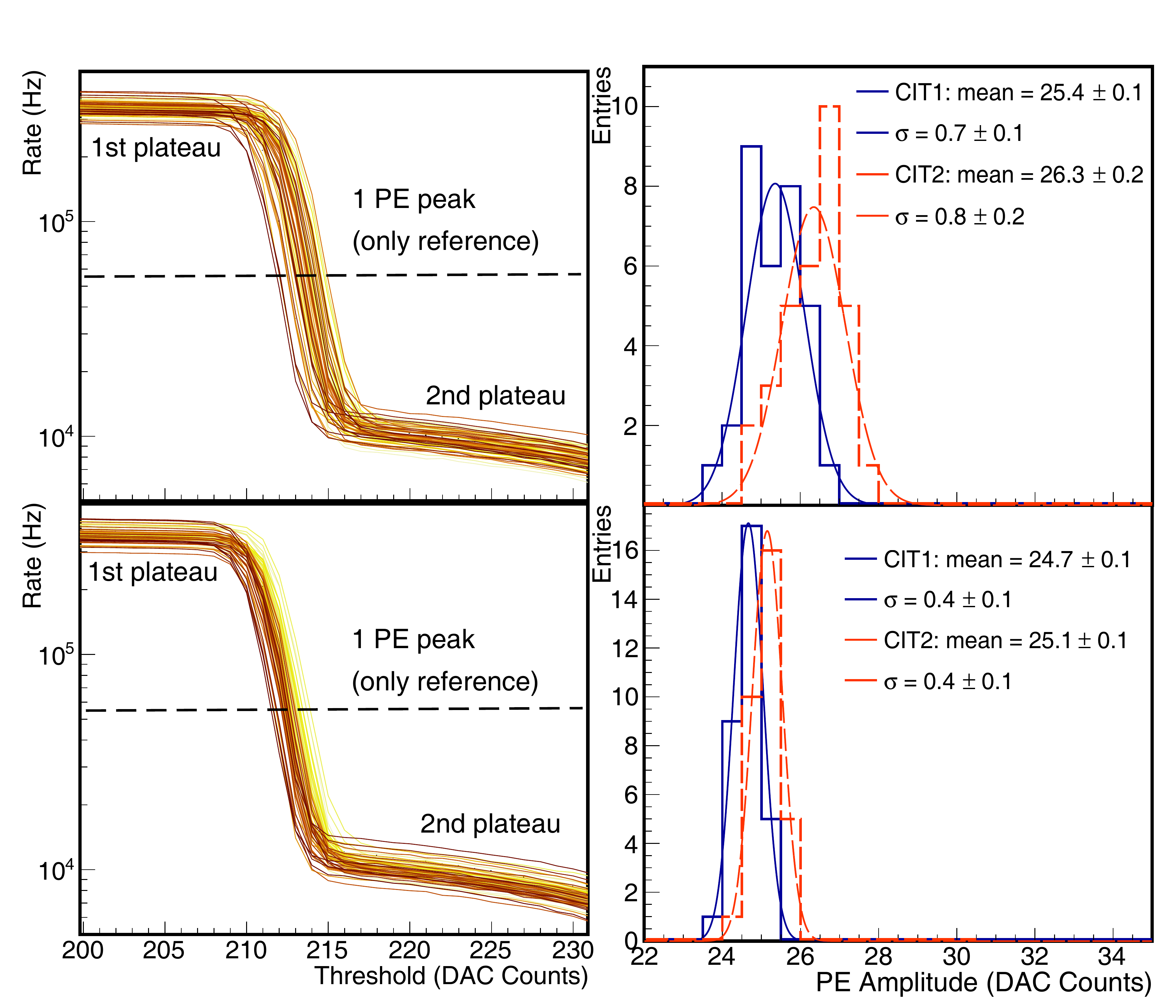}
\caption{Calibration: after breakdown voltages have been found, all SiPMs are set to the same over-voltage to assure uniform gain. The dark-rate curves and gain histograms are shown, before calibration (\textit{top}) and after calibration (\textit{bottom}).}
\label{fig::curves}       
\end{figure*}

A single silicon photomultipier is an array of photodiodes (referred as {\it micro-cells})
with densities that usually range from a hundred to several thousand per $\mathrm{mm^2}$. The photodiodes are operated with a reverse voltage ($V_\mathrm{bias}$) which exceeds the breakdown voltage ($V_\mathrm{br}$) of the silicon junction. 

$V_\mathrm{br}$ is the minimum voltage needed for SiPMs to operate in Geiger mode \cite{Renker}. If $V_\mathrm{bias} > V_\mathrm{br}$, each electron-hole pair created by the absorption of a photon produces a diverging avalanche of multiplication of carriers resulting in a macroscopic current flow in a single micro-cell. If the generated Geiger avalanche is not confined to the same micro-cell it was initiated in, crosstalk is produced, and the SiPM outputs a pulse
with twice (or more) the amplitude and charge of a single photo-equivalent (PE) pulse. As an illustration, in the left panel of fig.~\ref{fig::darkRate},
the oscilloscope capture of dark pulses of the AMIGA SiPM is shown. Therein, the discrete nature of the crosstalk pulses (as well as those of the afterpulses) is seen. Fig.~\ref{fig::calibration} depicts the PE amplitude spectrum arising from the crosstalk. The selected SiPM for AMIGA has 1584 micro-cells and, as already explained, there are 64 SiPMs per muon module coupled to the fibers through an optical connector 
(see fig.~\ref{fig::sipmarray}). It is known that SiPMs have a high dark rate so each of these 64 pixels might produce muon overcounting during data taking. Their dark rate (as well as crosstalk and gain) is proportional to the over-voltage ($\Delta V$), which is in turn determined by
$\Delta V = V_\mathrm{bias} - V_\mathrm{br}$. In the right panel of fig.~\ref{fig::darkRate}, the gain, dark rate, and crosstalk are shown as a function of $\Delta V$ while the experimental measurement of  $\Delta V$ is shown in the right-hand side of fig.~\ref{fig::calibration}.

The operating point of each AMIGA module is set by choosing a single $\Delta V$ value for all
its 64 SiPMs to assure a uniform gain for all of them. On top of this and regarding the counter-mode operation of the front-end electronics,
a discriminator threshold has to be selected. 
During the EA phase, these two parameters were optimized to enhance the detector performance. Note that increasing $\Delta V$ increases both noise and efficiency while a higher discrimination threshold reduces noise but may alongside reduce the detector efficiency.  Laboratory measurements \cite{Matias} have shown that both low background noise and high efficiency are achieved for $\Delta V = 3.5$\,V with a discriminator threshold set to $2.5~\mathrm{PE}$. A signal equal or above this threshold is counted as a muon in the UMD counter mode. This one-bit electronics makes muon identification fairly independent of fluctuations (light production, amplitude of PE signals, fiber attenuation, gain shifts).

\subsection{Calibration}
\label{sec::calibration}

Contrary to MPMTs, SiPMs have a well-defined PE spectrum. Therefore, it is possible to calibrate them using dark-rate pulses. The PE amplitude spectrum can be obtained by measuring the dark rate as a function of the discrimination threshold. The well-defined SiPM pulses result in turn
in well-defined plateaus as shown in the left panel of fig.~\ref{fig::calibration}. The transition from one plateau to the following earmarks the corresponding PE amplitude that can be readily identified via the curve derivative as also shown in the figure (green line). 
To set the $V_\mathrm{bias}$ (sec. \ref{sec::sipms}), it is necessary to first determine the $V_\mathrm{br}$ of each SiPM. 
To do this, dark-rate curves are measured at different $V_\mathrm{bias}$ as shown in the middle panel of fig.~\ref{fig::calibration}. It can be seen how the PE amplitude, dark rate and crosstalk (ratio between two plateaus) increase when $V_\mathrm{br}$ is increased. From these curves, the single PE amplitude as a function of the $V_\mathrm{bias}$ is obtained, as shown in the right panel of fig.~\ref{fig::calibration}. The extrapolation of this curve to 0 (in amplitude) indicates the voltage corresponding to $V_\mathrm{br}$. 

Once all 64 $V_\mathrm{br}$ are known for all 64 SiPMs, the $V_\mathrm{bias}$ is roughly adjusted with a Hamamatsu C11204-01 high-voltage source, which sets the same preliminary $V_\mathrm{bias}^{\mathrm{prel}}$ to all SiPMs. This value is determined by $V_\mathrm{bias}^{\mathrm{prel}} = \mathrm{max}\left\{ V_\mathrm{br} \right\} + \Delta V$, where $\mathrm{max}\left\{V_\mathrm{br} \right\}$ is the maximum value of the array of 64 SiPMs and $\Delta V$ is the desired over-voltage. 
To compensate for the $V_\mathrm{br}$ difference of each SiPM channel, a smooth adjustment per channel is performed with the ASIC which allows setting an individual voltage per channel to ensure the same $\Delta V$ for all SiPMs. 
As an example of the impact of the above described procedure,
the dark rate curves and the corresponding $1~\mathrm{PE}$ amplitude histograms are shown in fig.~\ref{fig::curves}, before (top) and after (bottom) the calibration run for $\Delta V = 3.5~\mathrm{V}$ (sec. \ref{sec::sipms}). It can be seen that after calibration the curves show less fluctuations, 
particularly the standard deviation in the $1~\mathrm{PE}$ amplitude 
is reduced by a factor of $\sim 2$ . Therefore the detector response becomes more uniform. 
\subsection{Temperature compensation}
\label{sec::temp}

\begin{figure} [htp]
\centering
\vspace*{-0.2in}
\includegraphics[width=8.5cm,clip]{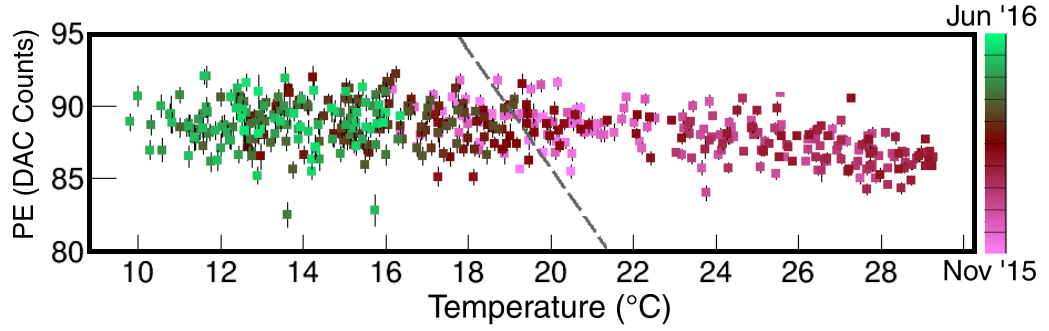}
	\caption{$1~\mathrm{PE}$ amplitude as a function of field temperature over an 8 month period with a temperature range of $\sim 10 - 30^\circ \mathrm{C}$. The colors indicate the months: greenish for the coldest season and reddish to the warmest. 
	The dotted-gray line shows the temperature dependence of the $1~\mathrm{PE}$ amplitude without any temperature compensation in the front-end electronics. As can be seen, the almost constant $1~\mathrm{PE}$ amplitude shows that the gain stabilization works at the level of $0.2\% / ^\circ \mathrm{C}$.
\label{fig::temperature}}
\end{figure}

The SiPM dark rate and its PE amplitude depend on temperature (see fig.~\ref{fig::temperature}). The dark rate is accounted for by the muon-counting strategy and the time duration of the air shower. In this section, we concentrate on characterizing the temperature impact on the PE amplitude.

\begin{figure} [htp]
\centering
\includegraphics[trim={0cm, 0cm, 0cm, 0.0cm},width=7.0cm,clip]{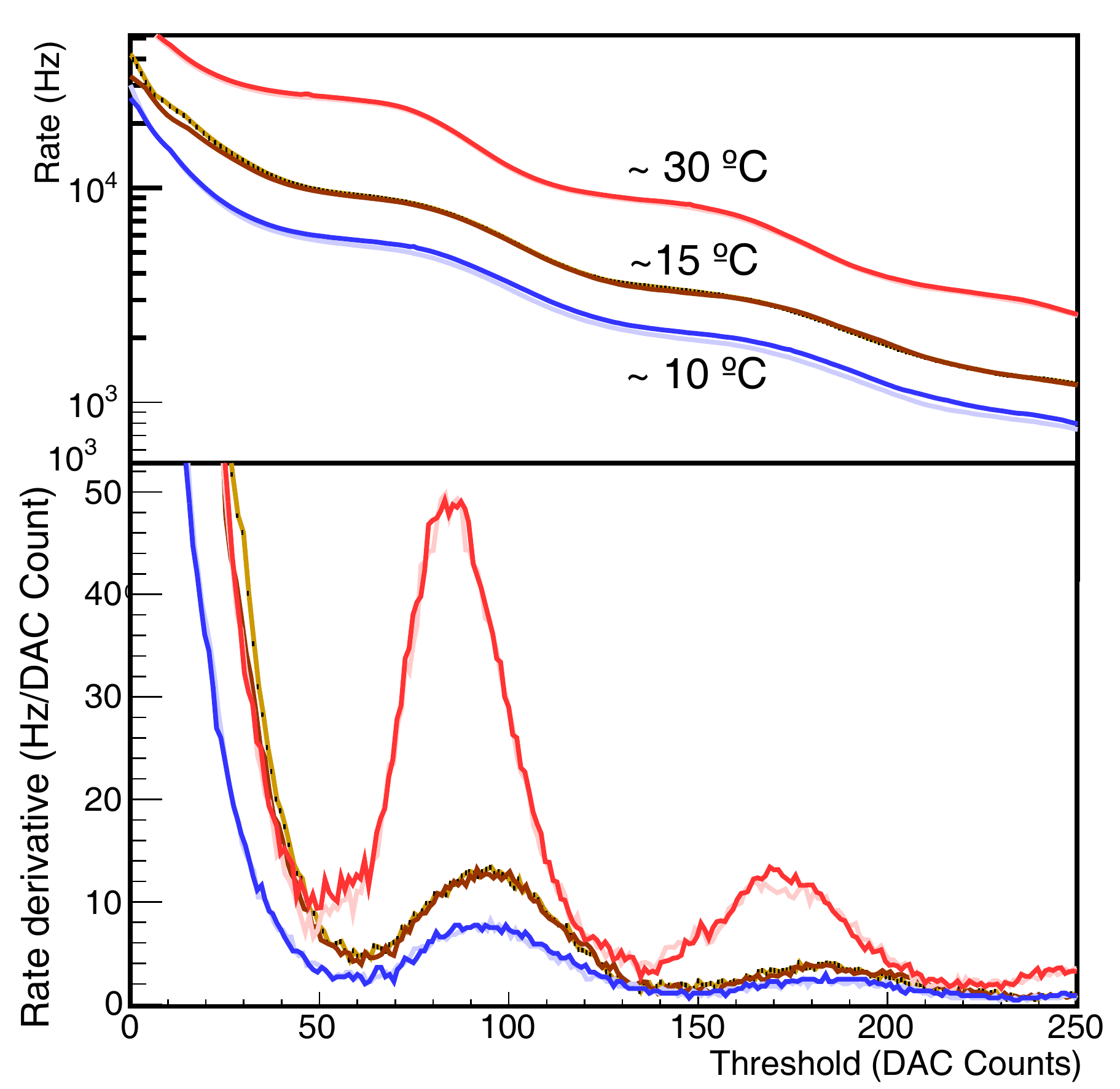}
\caption{Results for six different calibration runs at different temperatures. The shift in the $1~\mathrm{PE}$ spectrum can be observed. As a consequence of the temperature increase, the breakdown voltage increases accordingly and, therefore, the gain decreases.
\label{fig::tempDependance}}
\end{figure}

Dark-rate curves at different temperatures have been measured in the field with temperature sensors close to the SiPMs (see fig.~\ref{fig::tempDependance}). There is a noticeable shift in the $1~\mathrm{PE}$ amplitude due to temperature variations since, when temperature increases, the $\mathrm{V_{br}}$ increases accordingly and therefore $\mathrm{\Delta V}$ (along with the SiPM gain) decreases.

Field-temperature changes during the day and year and therefore it is crucial to provide automatic control of $V_\mathrm{bias}$ so as to maintain 
$\Delta V$ constant. By accomplishing this, the SiPM gain would remain stable over time. 
For the AMIGA detector modules, a temperature compensation mechanism is integrated into the Hamamatsu C11204-01 high voltage source. This mechanism adjusts $V_\mathrm{bias}$ according to the temperature variation thus compensating for any fluctuations in $V_\mathrm{br}$ and, therefore, stabilizing $\Delta V$. To study the performance of a compensation mechanism, 
the gain was estimated repeatedly for more than 8 months. Over periods lasting for approximately one to two weeks, both SiPMs gain and temperature were measured every two hours. The resulting dependence of gain on temperature is displayed in fig.~\ref{fig::temperature} where a temperature correspondence with seasonal changes can be observed. 
The data points acquired during the coldest months (May and June in the Southern hemisphere) cluster on the left-hand side while on the right-hand side, are those corresponding to the warmest months (January and February). Note that, since the UMC are buried 2.3 m deep, the overall temperature variation is only $\sim 20^\circ \mathrm{C}$.
The average gain over the entire period was $\mathrm{(88.7 \pm 1.6)}$ in DAC-counts units. The corresponding gain variation was approximately $\mathrm{0.2\% /^\circ C}$. Without the compensation, the variation would have been $\mathrm{4.3\% /^\circ C}$ \cite{datasheet2}. Thus, the compensation mechanism is reducing the gain variations by a factor of 20 (i.e. from $\sim 86$ to $4\%$).

\section{Muon reconstruction performance}
The two primary tasks of AMIGA UMDs are 
to count muons and to detect their arrival time. 
Muon counting can be performed individually muon-by-muon in the counter mode (one-bit electronics) and on a mean-value approach in the integrator mode (charge integration). The counter mode of the read-out electronics produces binary signals 
relying on the signal amplitude while the integrator mode is based on the total-signal charge. 

As SiPMs are located only at a single scintillator-strip edge, it is important to characterize the attenuation of the signal along the scintillator fibers and its impact on the muon-identification efficiency. In the following section,
the attenuation of both modes is quantified with laboratory measurements. For this purpose, a dedicated setup was mounted on a test-bench and the analog SiPM waveform pulses were directly acquired with an oscilloscope. The event trigger was performed by a small movable muon telescope and the whole length of the strips was swept in steps of 0.5\,m.

\subsection{Single-muon reconstruction}
\label{sec::attenuation}

\begin{figure} [H]
\centering
\includegraphics[width=7.0cm,clip]{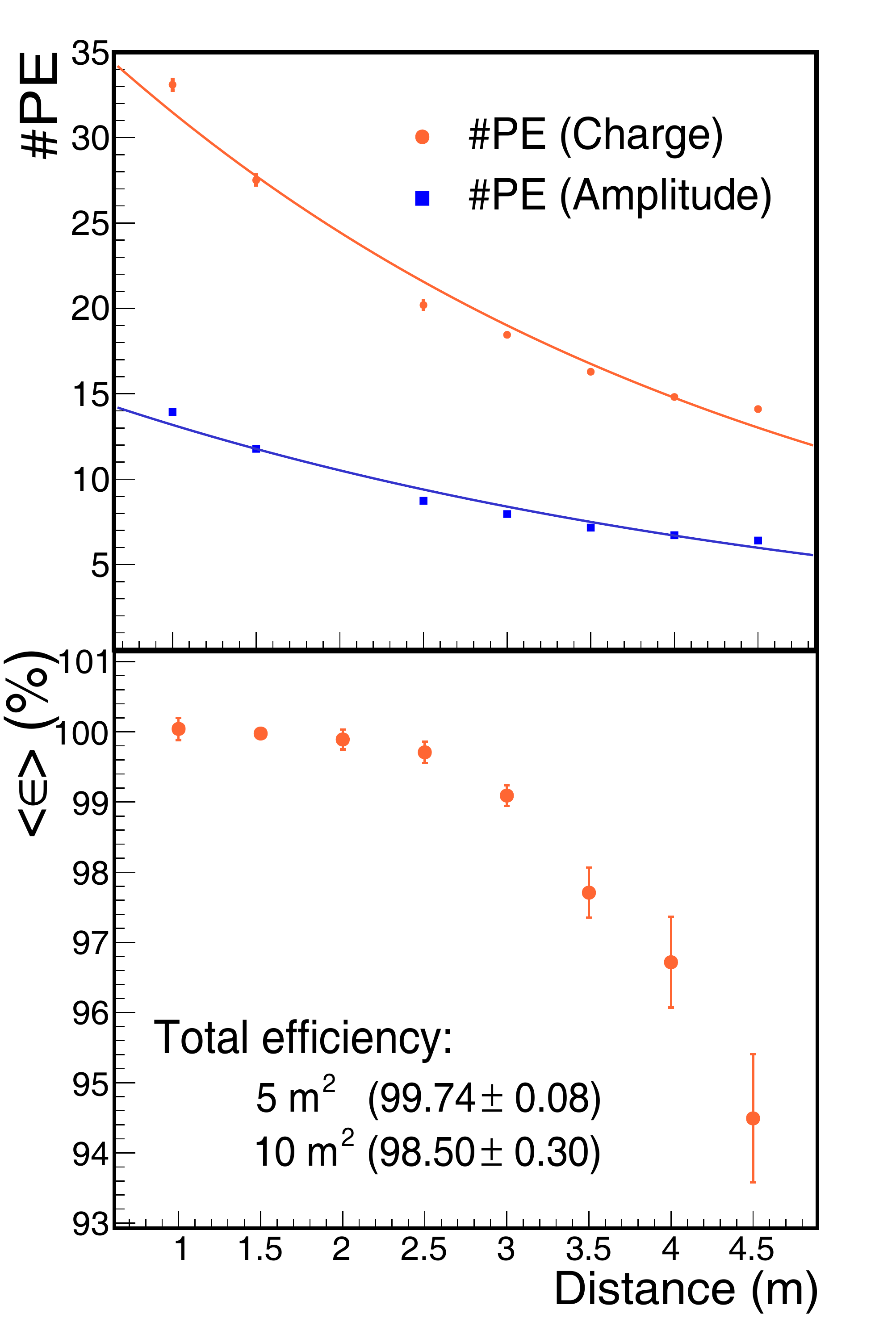}
	\caption{\textit{Top:} Number of PE as a function of the distance from the SiPM both in amplitude (sensitive to the {\it counter} mode) and in
	charge (sensitive to the integrator mode). \textit{Bottom: }Efficiency as a function of the distance from the SiPM for the counter mode. The total efficiency of the 5 and 10 $\mathrm{m^2}$ modules is estimated by integrating the efficiency curve between 1 m and 3 or 5 m  respectively. 
\label{fig::attenuation}}
\end{figure}

The number of PEs in the muon signals both in amplitude
and charge as a function of the distance to the SiPM is shown in the top panel of fig.~\ref{fig::attenuation}. The difference between amplitude and charge attenuations is related to the time distribution of photons reaching the optical sensor: the processes involved in the plastic scintillator as well as in the wavelength-shifter fiber spread photons in time and thus they do not add to the total SiPM pulse amplitude. Therefore, the final amplitude will not be the sum of the individual PE-signal amplitudes (they may be reduced by a factor $\approx 2$, see fig.~\ref{fig::attenuation} top) whereas the charge of the pulses will not be affected by this time spread since it is integrated over the total pulse timespan. Note that the counter mode is sensitive to the amplitude attenuation (as it implements an amplitude threshold) whereas the integrator mode is sensitive to the charge attenuation.
The impact of the attenuation on efficiency to detect muons in the
counter mode is shown in the bottom panel of fig.~\ref{fig::attenuation}. Typically, the SiPM dark rate (with 3 or more PE) surpasses the discriminator threshold for less than 12.5 ns. For this reason, muons are identified when the signal above the discriminator threshold lasts more than $12.5~\mathrm{ns}$. The efficiency loss at the end of the bar is the consequence of the PE  attenuation, as signals below the discriminator threshold ($2.5~\mathrm{PE}$) are lost. 
The total efficiency of the $5~\mathrm{m^2}$ and $10~\mathrm{m^2}$ modules is $99.7\%$ and $98.5\%$ respectively 
(the efficiency of the integrator mode has not yet been determined).

\subsection{Array reconstruction}
\label{sec::reconstruction}
\begin{figure} [H]
\centering
\vspace*{0.1in}
\includegraphics[width=7.0cm,clip]{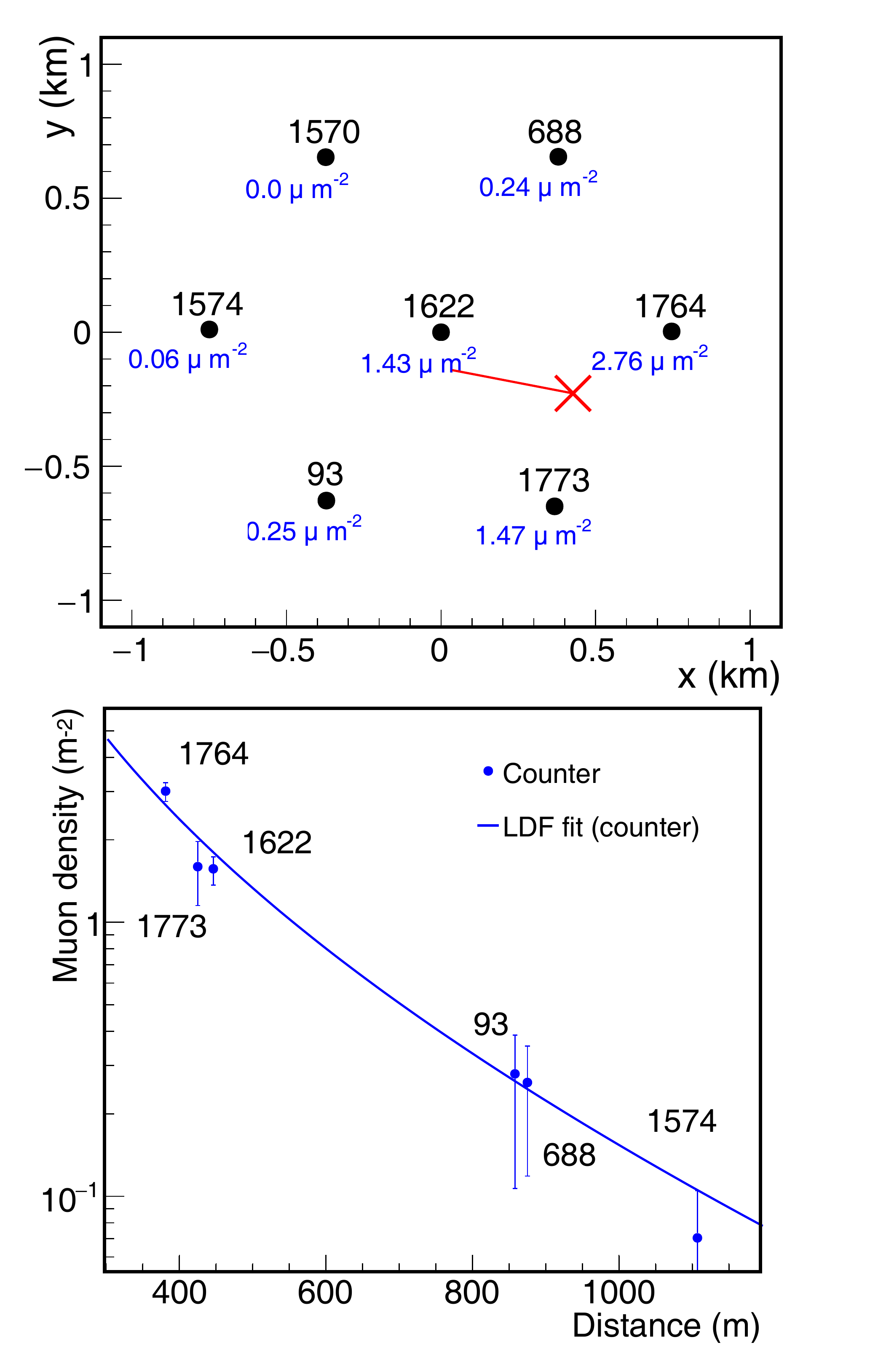}
	\caption{A real event reconstructed with the counter mode alone for an event with an energy of $\left(9.54 \pm 0.39 \pm 0.18 \right) \times 10^{17}$ eV and zenith angle of $\left(21.8 \pm 0.4\right)^\circ$. \textit{Top:} Footprint, core position, and azimuth angle. \textit{Bottom:} Lateral distribution of muon densities.
\label{fig::recEvent}}
\end{figure}

As an example of the performance of AMIGA as an array, the reconstruction of the muon lateral distribution \cite{reconstruction} in an air shower with reconstructed energy of $\left(9.54 \pm 0.39 \pm 0.18\right) \times 10^{17}~\mathrm{eV}$ and zenith angle of $\left(21.8 \pm 0.4\right)^\circ$ is shown in 
fig.~\ref{fig::recEvent}. Both air shower arrival geometry and energy were obtained from the surface WCD array, SD-750 \cite{reconstructionSD}.
Since the UMC integrator mode is yet to be fully calibrated, muon densities were reconstructed in the counter mode alone. The measured muon densities \cite{mueller} at each position are shown in the top panel of the figure while the fit to their lateral distribution is seen in the bottom panel.
The UMD array of AMIGA is expected to be completed by 2020. 

\section{Conclusion}
Preliminary results of the AMIGA engineering array have been analyzed and shown. A final satisfactory assessment in regard to both UMD construction (mechanics, telecommunications, optoelectronics, counter plus integrator electronics) and data analyses (calibration, temperature compensation, muon reconstruction performance) has been achieved. UMD production phase has started.

\end{document}